\documentclass[prl,superscriptaddress,a4paper,twocolumn,showpacs]{revtex4}
\usepackage{amsmath,graphicx}
\usepackage{mathptm}

\begin{document}

\title{The origin of stiffening in cross-linked semiflexible networks}
\author{P.~R.~Onck, T.~Koeman, T.~van Dillen, E.~van der Giessen}
\affiliation{Micromechanics of Materials, Materials Science
Centre, Nijenborgh 4, 9747 AG Groningen, The Netherlands}

%
%
\begin{abstract}
Strain stiffening of protein networks is explored by means of a finite
strain analysis of a two-dimensional network
model of cross-linked semiflexible filaments. The results show
that stiffening is caused by non-affine network rearrangements
that govern a transition from a bending dominated
response at small strains to a stretching dominated
response at large strains. Thermally-induced filament undulations
only have a minor effect; they merely postpone the transition.
\end{abstract}

 \maketitle


There is a deep interest in the mechanical response of biological
tissues and gels in view of the importance for biological
functions such as cell motility and mechanotransduction.  Many
network-like biological tissues respond to deformation by
exhibiting an increasing stiffness, i.e. ratio between change of
stress and change of strain. This has been demonstrated by
micropipette and microtwisting experiments \cite{Ingber} on
individual cells and through rheological experiments on in-vitro
gels of cytoskeletal filaments (actin, vimentin, keratin
\cite{Janmey90, Janmey91, ma} and neuronal intermediate filaments
\cite{Storm}), as well as on fibrin \cite{Hvidt, Shah}. These
biological gels fall within the class of \textit{semiflexible}
polymers, which has also attracted much theoretical
attention in the last decade \cite{MacKintosh, Isambert, Morse, Head, Frey03}.
However, these theoretical studies have primarily focused on the
small-strain regime, tractable for analytical treatment.

In a simple conceptual view, a biopolymer network is an
\textit{interlinked} structure of \textit{filaments}. Thus,
stiffening can result from stiffening of the polymeric filaments
between cross-links, from alterations in the network structure, or
both. The current paradigm is that stiffening is primarily due to
the stiffening of the filaments themselves. This idea has been
worked out in detail very recently by Storm \textit{et
al.}~\cite{Storm} by adopting the
worm-like chain model for actin filaments in combination
with the assumption that the network deforms in an affine manner,
i.e., each filament is assumed to follow the overall deformation.
The worm-like chain model is a well-documented description for the stretching
of semiflexible polymers, where the longitudinal stiffness of
undulated filaments is attributed primarily to bending; the axial
stiffness of the polymeric chain itself is much higher
\cite{MacKintosh,fn:entropy}. As the filament is stretched (at
constant temperature), the amplitude of the
transverse thermal undulations
reduces and, as a consequence, the stiffness increases. In the
limit that the filament is pulled straight, all subsequent axial
deformation would have to originate from axial straining of the
chain, but at an enormous energy cost. Given this description of
individual filaments, Storm \textit{et al.}~\cite{Storm} proceed by
considering a network consisting of infinitely many filaments.
Initially the filaments are randomly orientated, and as the sample
is deformed the network is assumed to distort in an affine manner.
The affine deformation assumption is well known in network models
for rubber elasticity, and allows for a relatively simple
description of the overall network response on the basis of the
behavior of a single filament. The small-strain affine deformation
assumption in two-dimensional networks of straight filaments has
recently been studied in great detail by Head \textit{et al.}~\cite{Head},
who conclude that its validity depends on the cross-link density
of the filaments. Their conclusion, however, cannot be immediately
transferred to the stiffening results of Storm \textit{et al.} since it
applies only to the initial response.

In contrast with the cited literature, we demonstrate in this
paper that stiffening lies in the network rather than in its
constituents. During deformation, the filaments rotate in the
direction of straining, which induces a transition from a
bending-dominated response to one that is controlled by stretching
of aligned filaments. By comparing cross-linked networks with and
without thermal undulations, we show that filament reorientation
is the dominant mechanism, while the thermal undulations only
postpone the onset of stiffening.


Our model is a two-dimensional network model of filaments in a
periodic unit cell of dimensions $W \times W$. The network is
generated by randomly placing filaments with length $L$ at random
orientations inside the cell, with proper account of periodicity.
Thermal undulations are mimicked by superposing on each filament
transverse
normal modes of the type $b_n \sin({n\pi x}/{L})$, where the
ampitudes $b_n$ follow a Gaussian distribution (cf.~\cite{Kas})
with standard deviation $\sqrt{2/(l_{\rm p} L)} (L/n\pi)^2$. The
persistence length $l_{\rm p}$, i.e., the distance over which the
filaments appear straight, is expressed as $l_{\rm p} \propto
{\kappa}/(k_{\rm B} T)$ in terms of the bending stiffness $\kappa$
of the filaments, Boltzmann's constant $k_{\rm B}$ and temperature
$T$. Clearly, the filaments are straight in case the temperature
is low or the bending stiffness is high.  We use the first 10
normal modes to generate the initial geometry of the filaments and
treat $l_{\rm p}$ as an independent quantity. During mechanical
loading, thermal effects are no longer taken into account.

Points where filaments overlap are considered to be cross-links,
similar to the procedure used by Head \textit{et al.}~\cite{Head}
and Wilhelm and Frey~\cite{Frey03}. The networks generated by this procedure
are taken as the initial, stress-free configuration. In the
calculations, the cross-links are assumed to be stiff, so that
both displacement and rotation of the two filaments at the
cross-link point remain the same. The filaments are elastic rods,
characterized by a stretching stiffness $\mu$ (axial force
\cite{fn:unit} needed to induce a unit axial strain) and a bending
stiffness $\kappa$ (bending moment needed to induce a unit radius
of curvature). For isotropic elastic rods these values are related
through their cross-sectional geometry, but are treated here as
independent. The density of the network is characterized by the
line density $\rho$, i.e., the total length of filaments in the
unit-cell divided by the cell area, $W^2$. For networks with
straight filaments ($l_{\rm p}/L \to \infty$), the average
distance between cross-links, $l_{\rm c}$, is inversely proportional to
$\rho$ through $l_{\rm c}=\pi/\rho$ \cite{Pike}.  We consider networks
of different densities, but only above the rigidity percolation
threshold \cite{Frey03}.

For the numerical study we use the finite element method,
discretizing each filament with 10 equal-sized Euler--Bernoulli
beam elements accounting for stretching and bending. Geometry
changes are accounted for by an updated Lagrangian finite strain
formulation. All filaments are perfectly bonded to rigid top and
bottom plates, with the top plate displaced horizontally relative
to the bottom plate over a distance $\Gamma W$, corresponding to
an applied shear strain $\Gamma$. The macroscopic shear stress
$\tau$ is calculated from the total horizontal reaction force at
the top, divided by $W$. Convergence studies ensured that the cell
size $W$ does not affect the results.

The parameters governing the
system are $\tau$, $\Gamma$, $\mu$, $\kappa$, $L$, $\rho$ and $l_{\rm p}$.
We choose to present the results through the following
dimensionless parameters: $\bar{\tau}=\tau L/\mu$, $\Gamma$,
$\bar{\rho}=\rho L$,
$\bar{l}_{\rm p}=l_{\rm p}/L$ and $\bar{l}_{\rm b}=\sqrt{\kappa/(\mu
L^2)}$.  Note that $\bar{l}_{\rm b}$ is a
measure for the 'floppyness' of the filaments, which reduces to
the slenderness ratio (thickness over length) for isotropic
elastic rods, and
$\bar{l}_{\rm p}$ sets the initial shape of the filaments.

\begin{figure}[htb]
\begin{center}
\includegraphics[width=\linewidth]{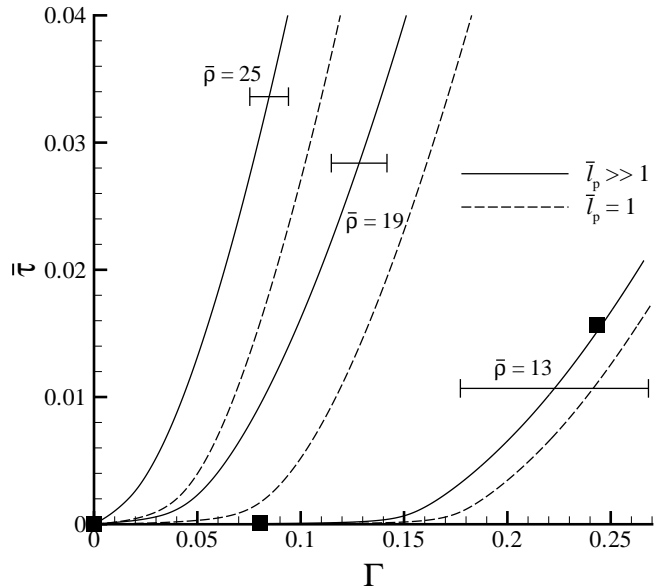}
\end{center}
\caption{Average stress ($\bar\tau$) versus strain ($\Gamma$)
response for networks with $\bar{l}_{\rm b}=2.3 \times 10^{-4}$
and $\mu/L=1.6$~MPa at three different densities $\bar\rho$ and
for straight ($\bar {l}_{\rm p} >>1$) and undulated ($\bar
{l}_{\rm p} = 1$) filaments. The error bars have a length of two
times the standard deviation in strain at a given $\Gamma$ for ten
different realizations at each density. The squares correspond to
three instances during deformation for which the network geometry
is shown in Fig.~\ref{fig2}.} \label{fig1}
\end{figure}
In a first set of calculations we take $\bar{l}_{\rm b}=2.3 \times
10^{-4}$ and $\mu/L=1.6$~MPa (representative for actin
microfilaments \cite{Howard}) with a density of $\bar{\rho}=13$,
which is well above the rigidity percolation threshold of
$\bar{\rho}=5.7$ \cite{Frey03}. The persistence length is taken to
be much larger than the filament length ($\bar{l}_{\rm p}
>> 1$), corresponding to
straight filaments. Figure~\ref{fig1} shows the stress-strain
response (averaged over ten different random realizations). Three
regimes can be identified: a regime with a relatively
low stiffness $d\bar\tau/d\Gamma$, a transition
regime and a high-stiffness regime.
\begin{figure*}[htb]
\begin{center}
\includegraphics[width=\linewidth]{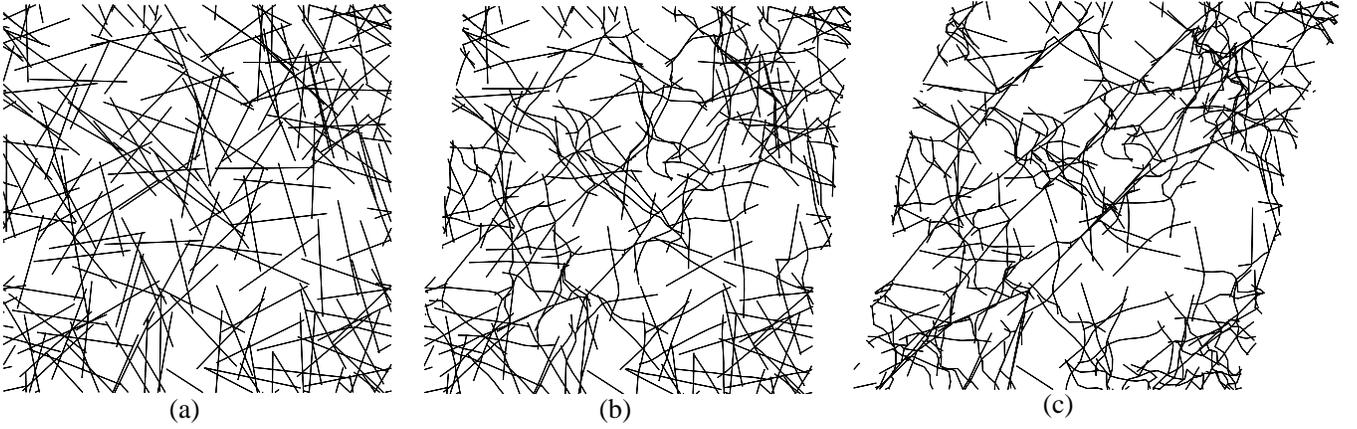}
\end{center}
\caption{(a) Initial, $\Gamma=0$; (b) intermediate, $\Gamma=0.08$
and (c) large strain, $\Gamma=0.24$, network configurations for a
typical $\bar{\rho}=13$ realization close to the average response
shown in Fig.~\ref{fig1} (squares).} \label{fig2}
\end{figure*}
Figure~\ref{fig2} shows three snapshots of the $\bar{\rho}=13$
network geometry at $\Gamma=0$, 0.08 and 0.24 for a typical
realization close to the average response (see the solid squares
in Fig.~\ref{fig1}). Comparison of Fig.~\ref{fig2}b with
Fig.~\ref{fig2}a reveals that many initially straight filaments
have deformed by bending, which corresponds to the characteristic
low stiffness at small strain levels for these densities
\cite{Head,Frey03}. Subsequently, during the transition regime,
percolations of stretched filaments appear that connect the top
and bottom of the cell along a $\sim 45^\circ$ direction,
Fig.~\ref{fig2}c. These filaments are loaded in axial tension,
resulting in a higher overall stiffness. Thus, Figs.~\ref{fig2}b
to \ref{fig2}c demonstrate the transition from a bending dominated
regime at small strains (total mechanical energy of the systems
primarily consists of bending energy) to a stretching dominated
regime at higher strain levels (total energy
dominated by the axial stretching energy).

\begin{figure}[h!]
\begin{center}
\includegraphics[width=\linewidth]{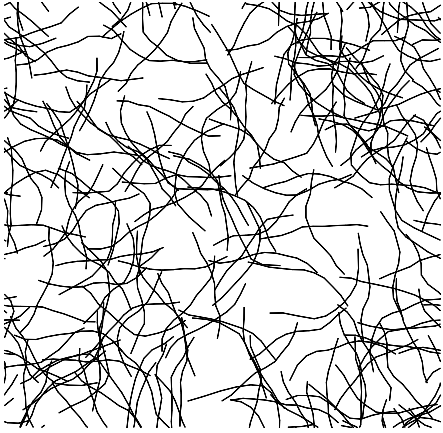}
\end{center}
\caption{Initial conformation of a $\bar{\rho}=13$ network with
undulated filaments corresponding to $\bar{l}_{\rm p}=1$.}
\label{fig3}
\end{figure}
The stress--strain response for two higher densities,
$\bar{\rho}=19$ and 25, is included in Fig.~\ref{fig1}. The
standard deviation in strain from ten realizations
is approximately independent of the
stress value for a given density. The scatter in strain, defined
as the ratio between the standard deviation and the average, is
independent of the density and has a value of approximately 0.2.
Figure~\ref{fig1} shows that a certain stress level is achieved at
smaller strains in case the network is denser. The network thus
gets stiffer with increasing density, while the transition from
bending to stretching becomes less abrupt and occurs at smaller
strain levels.

Next, the effect of thermally-induced undulations is investigated.
The same parameters are used as before, but now we use
$\bar{l}_{\rm p}=1$, in accordance with experimental findings for
actin filaments \cite{Howard}. Changing the persistence length
from $l_{\rm p} \gg L$ to $l_{\rm p}=L$ is physically similar to
increasing the temperature from $0$K to $293$K before
cross-linking and loading the network instantaneously.
Figure~\ref{fig3} depicts the initial geometry for a network where
the initial filaments' end-to-end vectors have the same location
and orientation as shown in Fig.~\ref{fig2}a. The stress--strain
results included in Fig.~\ref{fig1} (dashed lines) show that the
thermal undulations do not change the shape of the overall
stress--strain curve, but merely delay the transition from bending
to stretching. The associated `delay' strain at an applied strain
level of $\Gamma=0.25$ is 0.018, 0.029 and 0.024 for
$\bar\rho=13$, 19 and 25, respectively. Similar values were found for
densities up to 38. Clearly, only a small fraction of the total
strain is due to the presence of thermally-induced undulations.

\begin{figure}[htb]
\begin{center}
\includegraphics[width=\linewidth]{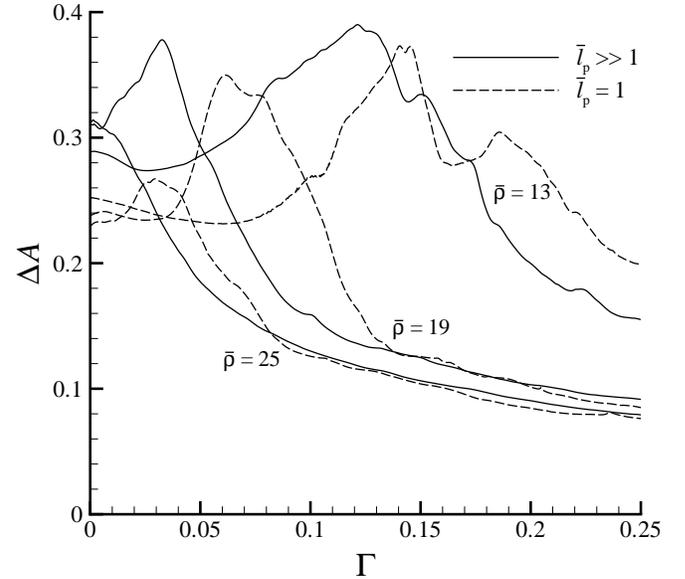}
\end{center}
\caption{The deviation from affine behavior, $\Delta{A}$, as a
function of strain, $\Gamma$, for $\bar{l}_{\rm b}=2.3 \times
10^{-4}$ and $\mu/L=1.6$~MPa, at three different densities and for
straight ($\bar{l}_{\rm p} >> 1$) and undulated ($\bar{l}_{\rm
p}=1$) filaments.} \label{fig4}
\end{figure}

To make connection with the small-strain study by Head et
 al.~\cite{Head}, we monitor the degree of affinity of the network
during straining to large deformations. For this purpose,
we define the deviation from affine behavior, $\Delta{A}$, as
\begin{equation}
\Delta{A}=\frac{1}{n} \sum_{k=1}^n \frac{ \| \Delta {\bf
r}^{(k)} - \Delta {\bf r}^{(k)}_{\rm aff} \| }{\Delta
\Gamma \| {\bf r}^{(k)} \| },
\end{equation}
where $\| {\bf r} \|^2={\bf r} \cdot {\bf r}$, $n$ the number of
cross-links and ${\bf r}^{(k)}$ is the current position vector of
cross-link $k$. $\Delta {\bf r}^{(k)}$ is the increment
in the position of cross-link $k$ during a shear increment $\Delta
\Gamma$ in the simulations, while $\Delta {\bf
r}^{(k)}_{\rm aff}$ is the corresponding value were the
deformation affine. Figure~\ref{fig4} shows the evolution of
$\Delta A$ as a function of $\Gamma$ for the cases shown in
Fig.~\ref{fig1}. It is observed that the deformation
is not affine at small strains, in accordance with
\cite{Head,Frey03}, while the deformation becomes increasingly
affine ($\Delta{A} \to 0$)
with increasing strain. By comparing Fig.~\ref{fig4} with
Fig.~\ref{fig1}, we find that in the transition from the bending
to the stretching regime, ${\Delta}A$ increases significantly,
indicating a reorientation of the filaments. The peak
in the ${\Delta}A$--$\Gamma$ curve occurs for both straight
and undulated filaments in the transition regime. This is
another indication of the fact that filament undulations do not
change the nature of network deformation, but merely enhance the
strain value at which stretching sets in. Once stretching has set in,
the deformation becomes more and more affine.

To study the influence of the filament properties, the
calculations are repeated, but with a larger bending and
stretching stiffness, $\bar{l}_{\rm b}=8.3 \times 10^{-4}$ and
$\mu/L=8$~MPa (representative for microtubuli \cite{MFvsMT}). Note
that the bending stiffness increases by a factor of 65, while the
stretching stiffness becomes five times larger. Because of the
enhanced bending stiffness, the persistence length at the high
temperature increases from $\bar{l}_{\rm p}=1$ to 65 so that the
filaments are almost straight (in accordance with experimental
observations on microtubuli \cite{Howard}).
\begin{figure}
\begin{center}
\includegraphics[width=\linewidth]{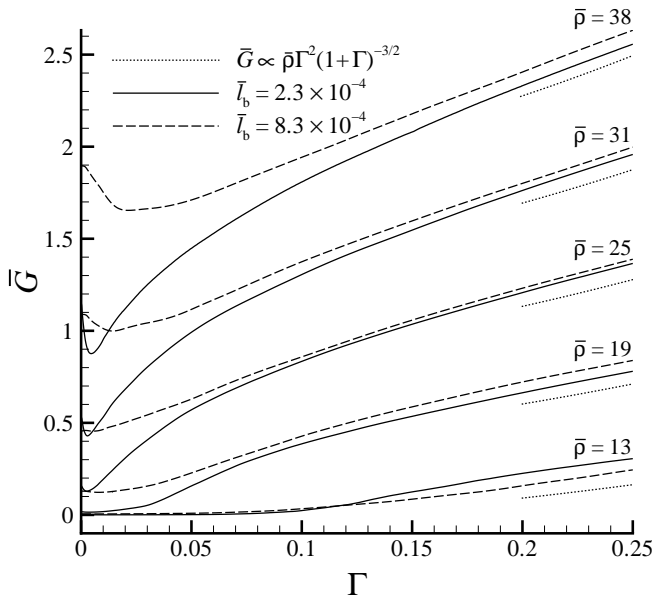}
\end{center}
\caption{Shear stiffness as a function of strain for networks with
straight filaments having $\bar{l}_{\rm b}=2.3 \times 10^{-4}$ or
$8.3 \times 10^{-4}$. The dotted lines represent the scaling
relationship (\ref{eq:finite-strain}), fitted to $\bar\rho=25$.}
\label{fig5}
\end{figure}
Figure~\ref{fig5} shows the instantaneous shear stiffness,
$\bar{G}=d\bar{\tau}/d\Gamma$, as a function of
shear strain $\Gamma$ for the two different values of
$\bar{l}_{\rm b}$ (and reference stress $\mu/L$). It can be
observed that for the `floppy' filaments ($\bar{l}_{\rm b}=2.3
\times 10^{-4}$) the transition from the low-stiffness to the
high-stiffness regime shifts to lower strains with increasing
density, consistent with Figs.~\ref{fig1} and \ref{fig4}. For
densities higher than $\bar{\rho}=25$, the transition from bending
to stretching is no longer accompanied by severe filament
reorientations that cause a peak in ${\Delta}A$ (see
Fig.~\ref{fig4}), but progresses more smoothly. At higher
densities of either floppy or stiff filaments, the stiffness first
decreases with strain at small strain levels. This is caused by
buckling of filaments oriented at $135^{\circ}$ away from the
horizontal axis (positive to the right in Fig.~\ref{fig2}), which
are loaded primarily in compression. Figure~\ref{fig5} also shows
that at small strains the overall stiffness for the floppy
filaments is much lower than that of the stiffer filaments, but
converges to the same value at larger strains. This  reflects that
bending is the dominant deformation mode in the low-stiffness
regime at small strains, while filament stretching governed by
$\mu$ dominates at large strains.

When filament stretching is the dominant deformation mechanism,
the following scaling arguments hold. The overall stress $\tau$ is
distributed, on average, over the filaments through the force $F
\propto \tau l_{\rm c}$, where $l_{\rm c} \propto 1/\rho$ is the
`mesh size' of the network. This force causes the filaments to
deform axially, yielding an elongation $\delta \propto F l_{\rm c}
/ \mu$, which yields an average strain $\Gamma \propto \delta
/l_{\rm c}$. Substitution yields $\Gamma \propto \tau / (\rho
\mu)$ so that $G \propto \rho \mu$. In three dimensions the same
scaling relation holds \cite{fn:3D}. It should be noted that in
the high-stiffness regime, the network `locks' when a number of
percolations connects top and bottom. Further straining the
network only results in stretching the (fixed number of)
percolations. In that case one would expect the stiffness to
converge to a fixed `steady state' value. However, due to
non-linear geometrical effects at large strains, the stiffness
increases with straining according to \cite{fn:non-linear geom}
\begin{equation}
\bar{G} \propto \bar{\rho} \Gamma^2 (1+\Gamma)^{-3/2}.
  \label{eq:finite-strain}
\end{equation}
This scaling is seen from Fig.~\ref{fig5} to successfully
characterize the steady state stretching regime at large strains.

This study leads to the following conclusions.
Stiffening of cross-linked semiflexible networks is caused by the
transition of a bending-dominated response at small strains to a
stretching-dominated response at large strains. This transition is
mediated by network rearrangements that are not affine. Filament
undulations only have a minor effect; they merely postpone the
transition from bending to stretching. Above a density-dependent
transition strain, the network stiffness scales linearly with
density and the filament's stretching stiffness, which is expected
to hold in three dimensions as well.


\begin{thebibliography}{9}
\bibitem{fn:entropy} Bending energy is internal energy, but since the
bending inside a filament is induced by transverse thermal undulations,
the stiffness according to the worm-like chain model is often referred to as
`entropic stiffness'. It should, however, be distinguished from
the classical entropy elasticity in flexible polymers.
\bibitem{fn:unit} All quantities with dimension force are computed
per unit out-of-plane thickness.
\bibitem{MFvsMT} Consider a microtubule (MT) having
an inner radius of 8 nm and an outer radius of 12 nm. If we
compare the stretching and bending stiffness with that of an actin
microfilament (MF), represented by a solid rod of radius 4 nm, it
follows that $\mu_{\rm MT}/\mu_{\rm MF}=5$, $\kappa_{\rm
MT}/\kappa_{\rm MF}=65$.
\bibitem{fn:3D} In three dimensions we have ${F} \propto \tau l_{\rm
c}^2$ and $l_{\rm c} \propto 1/\sqrt{{\rho}}$, leading again
to $G \propto {\rho} {\mu}$.
\bibitem{fn:non-linear geom} Consider a `network' consisting of
parallel percolated filament strings that are initially normal to
the shear plane and a distance $\propto 1/\rho$ apart. The strings
only have an axial stiffness $\mu$. Upon application of a shear
strain, the strings only rotate initially, but when straining
proceeds and the strings rotate by an angle $\beta =
\tan^{-1}\Gamma$, they have to stretch, thus leading to an
increase of the overall stiffness. Straightforward mechanics
yields $G \propto \rho \mu \cos{\beta} \sin^2\beta = \rho \mu
\Gamma^2 (1+\Gamma)^{-3/2}$, leading to
Eq.~(\ref{eq:finite-strain}). The same scaling holds in three
dimensions.
\end{thebibliography}
\end{document}